\documentclass[conference]{IEEEtran}
\IEEEoverridecommandlockouts
\usepackage{cite}
\usepackage{amsmath,amssymb,amsfonts}
\usepackage{algorithmic}
\usepackage{graphicx}
\usepackage{textcomp}
\usepackage{algorithm,algorithmic}
\usepackage{mathtools}
\usepackage{graphicx,subfigure}
\usepackage{ragged2e}
\usepackage{siunitx} 
\usepackage{dblfloatfix}

\DeclareMathOperator*{\argmin}{arg\,min}

\usepackage{xcolor}
\def\BibTeX{{\rm B\kern-.05em{\sc i\kern-.025em b}\kern-.08em
    T\kern-.1667em\lower.7ex\hbox{E}\kern-.125emX}}
\begin{document}

\onecolumn
\textbf{GOVERNMENT LICENSE}

The submitted manuscript has been created by UChicago Argonne, LLC, Operator of Argonne
National Laboratory (``Argonne''). Argonne, a U.S. Department of Energy Office of Science laboratory, is operated under Contract No. DE-AC02-06CH11357. The U.S. Government retains for
itself, and others acting on its behalf, a paid-up nonexclusive, irrevocable worldwide license in
said article to reproduce, prepare derivative works, distribute copies to the public, and perform
publicly and display publicly, by or on behalf of the Government. The Department of Energy will
provide public access to these results of federally sponsored research in accordance with the DOE
Public Access Plan. http://energy.gov/downloads/doe-public-access-plan.
\newpage
\twocolumn

\title{Optimizing Paths for Adaptive Fly-Scan Microscopy: An Extended Version}

\author{
  Yu Lu$^{\star,\dagger}$ \qquad
  Thomas F. Lynn$^{\dagger}$ \qquad
  Ming Du$^{\dagger}$ \qquad
  Zichao Di$^{\dagger}$ \qquad
  Sven Leyffer$^{\dagger}$ 
  
  \medskip 
  \medskip
  \\
  
  $^{\star}$University of California, Merced, CA, USA \\
  $^{\dagger}$Argonne National Laboratory, Lemont, IL, USA
}

\maketitle

\begin{abstract}
    In x-ray microscopy, traditional raster-scanning techniques are used to acquire a microscopic image in a series of step-scans. Alternatively, scanning the x-ray probe along a continuous path, called a fly-scan, reduces scan time and increases scan efficiency. However, in many imaging scenarios, not all regions of an image are equally important. Currently used fly-scan methods do not adapt to the characteristics of the sample during the scan, often wasting time in uniform, uninteresting regions. While we are not aware of any adaptive solution for fly-scan methods, one approach to avoid unnecessary scanning in uniform regions for raster step-scans is to use deep learning techniques to select a shorter optimal scan path instead of a traditional raster scan path, followed by reconstructing the entire image from the partially scanned data. However, this approach heavily depends on the quality of the initial sampling, requires a large dataset for training, and incurs high computational costs. We propose leveraging the fly-scan method along an optimal scanning path, focusing on regions of interest (ROIs) and using image completion techniques to reconstruct details in non-scanned areas. This approach further shortens the scanning process and potentially decreases x-ray exposure dose while maintaining high-quality and detailed information in critical regions. To achieve this, we introduce a multi-iteration fly-scan framework that adapts to the scanned image. Specifically, in each iteration, we define two key functions: (1) a score function to generate initial anchor points and identify potential ROIs, and (2) an objective function to optimize the anchor points for convergence to an optimal set. Using these anchor points, we compute the shortest scanning path between optimized anchor points, perform the fly-scan, and subsequently apply image completion based on the acquired information in preparation for the next scan iteration.
\end{abstract}


\section{Introduction}
\label{sec:intro}

Scanning microscopy is an imaging technique where a focused beam (of x-rays or electrons, for example) is scanned across the surface of the sample being measured. At each scanned location, a detector collects the measured intensity of the emitted signal from the sample. Conventionally, this technique is conducted in as a series of "step-scans": for each scanned point, the sample stage needs to come to a full stop before signals are collected while the sample is stationary. The stage then accelerates from static and moves to the next point. Compared to the traditional step-scan, a new method called "fly-scan" has been introduced and has been popular for its high-speed scanning capabilities \cite{clark2014continuous, pelz2014fly, deng2015continuous}.
In fly-scan, the sample stage moves constantly, and signals are collected as the sample is in motion. Although this introduces motion blur, ultra-bright light sources, such as synchrotron facilities that generate high-flux x-ray beams, can yield sufficient signals within a short exposure window, thereby keeping the motion blur well under control. 

\begin{figure}[H]
    \centering
    \includegraphics[width=0.98\linewidth]{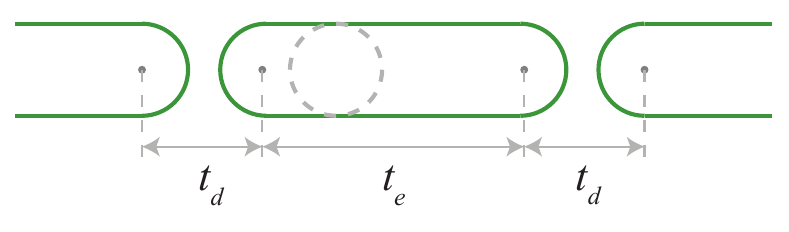}
    \caption{A schematic diagram illustrating the key components of fly-scan microscopy. Assuming the probe has a circular footprint on the sample (gray dashed circle), it sweeps the area enclosed by the green line during each exposure time $t_e
    $. The signals integrated during this window constitute a detector readout. Between exposure windows, there is a short dead time $t_d$ where the detector processes and sends out the measured value, during which no signal is collected.}
    \label{fig:flyscan_scheme}
\end{figure}

In fly-scan, four critical concepts are central to its operation: exposure time, dead time, scan speed, and beam size. The exposure time is the period during which the detector actively collects data; the detector reports the integrated signal intensity during each exposure window as a readout value. The dead time is the time when no data is collected due to processing or hardware limitations. With state-of-the-art techniques, the dead time is typically much shorter compared to the exposure time \cite{wahl2020photon}, enhancing the efficiency of data acquisition. The scan speed is the velocity at which the scanning system moves across the sample. The speed directly affects spatial resolution and integration time. Faster motion reduces the sampling time at each point, but can potentially compromise scan quality. Lastly, the beam size is the dimensions of the probe or beam (e.g., X-ray or laser) used to interact with the sample. Fig.~\ref{fig:flyscan_scheme} illustrates these key components of fly-scan.

In many imaging scenarios, not all regions of an image are equally important. Using a raster fly-scan path, where the probe or sample stage moves in a regular pattern so that the probe visits the entire sample surface, often results in unnecessary scanning of uniform regions. To address this, this paper proposes an adaptive fly-scan framework that prioritizes scanning regions of interest (ROIs) along a specific designed optimal scan path, while employing image completion techniques to reconstruct details in non-scanned areas. Using this proposed technique, the scan time can be reduced to one-third of the original time, while maintaining high-quality final images after image completion. Specifically, the main contributions of our work are summarized as follows:
\begin{itemize} 
    \item We propose an adaptive fly-scan framework that utilizes less than $30\%$ of the measurements in dense scan while maintaining high-quality final images through an image completion technique.
    \item We define an efficient score function and objective function to generate optimal anchor points for scan routing. 
    \item We design three experiments to verify the effectiveness of the proposed functions and demonstrate the generalization capability of the proposed framework.
\end{itemize}

\section{Related Works}
Beyond the basic concepts of fly-scan, it is important to review adaptive scanning and image completion techniques.

\subsection{Adaptive Scanning}
In general, scanning approaches can be categorized into raster scanning \cite{negroponte1977raster} and adaptive scanning \cite{shi2015novel}, depending on whether the full image or only a portion of it is scanned. Additionally, scans can be classified as fly-scan \cite{huang2015fly} or step-scan \cite{muller2010image} based on whether the scanning motion is continuous or discrete.

Raster scanning is the baseline approach widely used in research. Some variant methods of raster scanning use space-filling curves, instead of a rectangular scan grid, to maximize the coverage of the probe in the sample surface, such as specially designed raster scan paths \cite{odstrvcil2018arbitrary} or varying scan speeds across different areas of the image \cite{wang2019scan}. Adaptive scanning was later introduced for use with step-scan techniques \cite{shi2015novel}. More recently, deep learning methods have been applied to adaptive step-scan frameworks \cite{du2024predicting}, achieving impressive results \cite{kandel2023demonstration}. However, to the best of our knowledge, these adaptive scanning approaches have not yet been extended to fly-scan. This serves as one motivation for exploring the proposed adaptive fly-scan framework. However, considering the dataset requirements and training time, we propose an analytical adaptive fly-scan framework rather than employing deep learning-based methods.

\subsection{Image Completion}
Adaptive scanning produces only partial information about the target image. To recover information about the complete image, matrix completion or image completion is used. Many image completion algorithms have been proposed \cite{cai2010singular, candes2012exact, bertsimas2020fast}. Some advanced algorithms can simultaneously perform denoising and reconstruction \cite{candes2010matrix, cao2015poisson, lu2024negative}.

However, these algorithms often require longer computation times and are designed to handle arbitrary missing information. In our framework, the fly-scan focuses on important, feature-rich regions and avoids scanning featureless areas. In other words, the available information is already key and feature-dense, making our image completion much simpler. Considering our goal of enhancing the efficiency of the fly-scan process, we select an inverse distance-weighted (IDW) interpolation method \cite{lu2008adaptive} as our completion approach.

\section{Problem Formulation}

We aim to reconstruct the ground truth image $f^* \in \mathbb{R}^{m \times n}$ approximated by an iteration of scans as  
\begin{align}
    f^{k+1} = \mathcal{R}(\mathcal{F}(\Gamma_k^*))
\end{align}
where the path $\Gamma_k^{*}$ is the optimal path in the $k$-th iteration, function $\mathcal{F}$ represents the fly-scan to obtain the image information along the path, $f^{k+1}$ is the $(k+1)$-th reconstruction, and function $\mathcal{R}$ is the IDW interpolation method. The explicit formulation of the IDW method is discussed in the next section.

Optimizing the path $\Gamma_k^*$ between the optimal anchor points set $\omega_k^*$ in the $k$-th iteration requires considering the previous $k-1$ scans. The combined anchor points $\Omega_k$ and combined optimal anchor points $\Omega_k^*$ can be represented as
\begin{align*}
    \Omega_{k} &= \omega^*_0 \cup \cdots \cup \omega^*_{k-1} \cup \omega_{k},\\
    \Omega_{k}^* &= \omega^*_0 \cup \cdots \cup \omega^*_{k-1} \cup \omega^*_{k}
\end{align*}
where the $\omega^*_k$ is the set of new additional anchor points at the $k$-th iteration. A direct approach for finding $\Gamma_k^*$ is to solve the Traveling Salesman Problem (TSP) to obtain the exact shortest path between anchor points $\Omega^*_k$. However, solving the TSP is computationally expensive, and the true shortest path may not be optimal for our objective function, as will be discussed further in the next section. An alternative approach is to use a heuristic algorithm to approximate the TSP shortest path more efficiently. This is denoted as  
\begin{align}\label{eq:huristic}
    \Gamma_k^* = \mathcal{H}(\Omega^*_k)
\end{align}
where $\mathcal{H}$ is the heuristic algorithm used.

We obtain the set of new optimal anchor points $\Omega_k^*$ using previous anchor points $\Omega_{k-1}^*$ and the current image reconstruction $f^k$, which can be stated as  
\begin{align}
    \Omega^*_k = \argmin_{\omega_{k}} \mathcal{L}(\Omega^*_{k-1} \cup \omega_k, f^k) = \argmin_{\Omega_{k}} \mathcal{L}(\Omega_k, f^k)
\end{align}
where $\mathcal{L}$ is the objective function, and $\Omega_k = \Omega^*_{k-1} \cup \omega_k$. We use a gradient-based algorithm to solve this optimization problem starting from an initial set of anchor points $\Omega_k^0$, which is computed as  
\begin{align}\label{eq:score}
    \Omega_k^0 = \mathcal{G}(f^{k}, N)
\end{align}
where $N$ is the size of the initial anchor points set $\Omega_k^0$ for the $k$-th iteration and $\mathcal{G}$ is the proposed score function to generate the initial anchor points based on the previous reconstruction.

\section{Method}
From the last section, the proposed adaptive fly-scan framework requires solving a 4-step problem. In this section, we detail solutions to each sub-problem.

\subsection{Objective Function}

The primary goal of the optimization is to generate a new set of points that minimizes the distance between the reconstruction and the ground truth. Specifically, the goal is
\begin{align*}
    \| \mathcal{R}(\mathcal{F}(\mathcal{H}(\Omega^*_k))) - f^*\|_F \leq \| \mathcal{R}(\mathcal{F}(\mathcal{H}(\Omega_k))) - f^*\|_F,
\end{align*}
where $F$ denotes the Frobenius norm. However, since the ground truth image $f^*$ is unknown, we require a loss function $\mathcal{L}$ that approximates the above such that
\begin{align*}
    \Omega^*_k = \argmin_{\Omega_k} \mathcal{L}(\Omega_k, f^k) \approx \argmin_{\Omega_k} \| \mathcal{R}(\mathcal{F}(H(\Omega_k))) - f^*\|_F.
\end{align*}
The proposed objective function in our framework is
\begin{align}\label{EQ:AC}
    & \mathcal{L}(f^k, \Omega_k) = - ({\alpha \log \mathcal{U}(\Omega_k) + \log \| \nabla \mathcal{I}(f^k, \Omega_k) \|_2})
\end{align}
where $\alpha$ is a constant, the first term is proposed exponentially weighted uncertainty function (EWUF) representing uncertainty, and the second term represents the gradient of the potential new reconstruction. The EWUF is formulated as
\begin{align}\label{eq:EWUF}
    \mathcal{U}( \Omega_k) =  \frac{1}{N} \sum_{i=1}^N \left ( \sigma^2 \left( 1 - \exp \left( - \sum_{j=1}^{M_i} \lambda_{i,j} d_{i,j} \right) \right) \right ),
\end{align}
where
\begin{itemize}
    \item $d_{i,j} = \frac{\|\mathbf{\xi}_i - p_j\|^2}{\ell^2}$ is the normalized squared distance between the $i$\nobreakdash-th query point $\mathbf{\xi}_i = (x_i, y_i) \in \omega_k$ and the $j$\nobreakdash-th reconstruction point $p_j = (x_j, y_j)$ corresponding to $\mathbf{\xi}_i$, where $p_j$ is one of the $M_i$ nearest neighbors to $\xi_i$ in $\Omega^*_{k-1}$,
    \item $\lambda = \frac{\exp(-d_{i,j})}{\sum_{i,j}^N \exp(-d_{i,j})}$ is the softmax weight for $d_{i,j}$,
    \item $\ell$ is the characteristic length scale,
    \item $\sigma$ is the noise standard deviation, which a estimating of the standard deviation of image noise. Larger noise leads to higher uncertainty. 
\end{itemize}
The gradient term $\nabla \mathcal{I}$ in \eqref{EQ:AC} is critical because high-gradient areas in an image typically contain more information and are more important. This norm of this gradient is computed as
\begin{align}
    \| \nabla \mathcal{I}(f^k, \Omega_k) \|_2 = \sqrt{\left( \frac{\partial f^k}{\partial x} \right)^2 + \left( \frac{\partial f^k}{\partial y} \right)^2}.
\end{align}
When an image is viewed as discrete, we use the central difference method to approximate its gradient:
\[
\frac{\partial f^k}{\partial x} \approx 
\frac{f^k[i, j+1] - f^k[i, j-1]}{2},
\]
\[
\frac{\partial f^k}{\partial y} \approx 
\frac{f^k[i+1, j] - f^k[i-1, j]}{2}.
\]

The motivation of \eqref{EQ:AC} is very straightforward. It is designed to address the problem that we do not have access to the ground truth, making it impossible to directly minimize the distance between the reconstruction and the ground truth. The first term is the exponentially weighted uncertainty function, which estimates the uncertainty of the potential new reconstruction, while the second term represents the gradient of the potential new reconstruction. The purpose of having both uncertainty and gradient terms in the objective function is to balance exploration and exploitation: the uncertainty term prompts the algorithm to explore previously under-sampled regions to avoid missing important features not yet detected, and the gradient term drives the algorithm to exploit fast varying regions containing high-frequency features that require additional resolution.

With the well-defined objective function in eq.~\ref{EQ:AC}, we can solve for \( \Omega_k^* \) using a gradient-based algorithm as follows
\begin{align}\label{eq:ADAM}
    \Omega_k^{s+1} = \Omega_k^{s} - \beta \nabla_{\Omega^s_k} \mathcal{L}(f^k, \Omega^s_k)
\end{align}
which is equivalent to
\begin{align*}
    (x^{s+1}_i, y^{s+1}_i) = (x^{s}_i, y^{s}_i)  - \beta \nabla_{(x^{s}_i, y^{s}_i)} \mathcal{L}(f^k, \Omega_k^s)
\end{align*}
where $\beta$  is the learning rate used in the ADAM optimizer \cite{kingma2014adam} and points $(x^{s}_i, y^{s}_i) \in \omega_k^s$.

By minimizing the objective function, we select the anchor points set that has the highest combined uncertainty and importance, inspired by acquisition functions typically used for Bayesian optimization \cite{kamperis2021acquisition}. Conducting the fly-scan along the path consisted by these anchor points provides ground truth information for those regions. Consequently, incorporating these points into the reconstruction significantly improves accuracy.

\subsection{Score Function}
In theory, if the combination of objective function and optimizer converge to the limiting points quickly and accurately, minimizing the distance between the reconstruction and the ground truth, we can start with any initial points $\omega_k^0$ and still achieve very good results. However, since we use ADAM, which converges to a local minimizer, and an approximation function as the loss function, convergence can be slow and the results may not be accurate. Therefore, we propose a score function to more intelligently select the initial points at each iteration, improving convergence and accuracy.

Since our loss function aims to select new anchor points that reflect the highest uncertainty and importance, the score function can be designed to strengthen these two aspects. In this paper, we propose a score function based directly on the gradient. Specifically, we define the score function \( \mathcal{G}(f^k, N) \) to select $N$ points from the distribution
\begin{align}
    P(x_i, y_i) = 
    \begin{cases} 
        \frac{ | \nabla_{(x_i, y_i)} f^k |}{ \sum_{j} | \nabla_{(x_j, y_j)} f^k |}, & \text{if } \| \nabla f^k \| > 0, \\
        0, & \text{otherwise}.
    \end{cases}
\end{align}
Here, \( \mathcal{G}(f^k, N) \) selects the \( N \) anchor points with the highest probabilities, and the index $j \in J$ is the set of all pixels in the reconstruction.

\subsection{Heuristic Algorithm for Path Generation}
After using the score function to generate the initial anchor points \( \Omega_k^0 \) and applying the objective function with a gradient-based algorithm, we obtain an optimal anchor points set \( \Omega_k^S \approx \Omega_k^* \). Based on the anchor points set \( \Omega_k^* \), it is theoretically possible to generate the shortest path by solving the TSP problem \cite{junger1995traveling, hoffman2013traveling}. However, the computational cost for this approach is very high. Therefore, we apply a nearest neighbor algorithm \cite{hu2006fast, larose2014k, sun2010adaptive} to generate an approximate shortest path.

Formally, given a set of points, the algorithm constructs a path by iteratively selecting the closest unvisited point to the current location until all points are visited. For a given set of anchor points $\Omega = \{\omega_1, \omega_2, ..., \omega_N\}$, starting from an initial point $\omega_0^*$, the next point $\omega_{i+1}^*$ is selected as
\begin{align}\label{eq:HA}
\omega_{i+1}^* = \argmin_{\omega \in \chi \subseteq \Omega \setminus \{\omega_0^*, ..., \omega_i^*\}} \| \omega_i - \omega \|_2,
\end{align}
where $\chi$ is a subset of candidate anchor points and $\| \cdot \|_2$ denotes the Euclidean distance. The size of $\chi$ determines the complexity of the heuristic algorithm. If $\chi$ equals the entire set $\Omega \setminus \{\omega_0^*, ..., \omega_i^*\}$, the procedure reduces to the standard greedy nearest neighbor algorithm for TSP problem. In our paper, this process repeats until all points are visited, producing an approximate path, \( \Gamma_k^* \), that connects all points in $\Omega$.

\subsection{Image Completion by IDW}
The fly-scan along the path \( \Gamma_k^* \), generated by the nearest neighbor algorithm, obtains ground truth values along the path. This leads our final subproblem to a typical image completion problem. Many fast and accurate methods have been developed for this task \cite{sun2005image, iizuka2017globally, drori2003fragment}. In this paper, we extend the available information to the rest of the image using the inverse distance weighting (IDW) method for matrix completion \cite{kandel2023demonstration, zhang2024optimized}:
\begin{align}\label{eq:IDW}
    \mathcal{R}(\mathcal{F}(\Gamma_k^*)) = \frac{\int_{\Gamma_k^*} W(\mathbf{p}) \mathcal{F}(\mathbf{p}) \, d\mathbf{p}}{\int_{\Gamma_k^*} W(\mathbf{p}) \, d\mathbf{p}},
\end{align}
where
\begin{itemize}
    \item \( W(\mathbf{p}) \) is the weight function, defined as \( W(\mathbf{p}) = \frac{1}{\|\mathbf{\sigma} - \mathbf{p}\|_2^2} \), where the query location \( \mathbf{\sigma} \) is a location not scanned, and \( \mathbf{p} \in \Gamma_k^* \) is a location that has been fly-scanned,
    \item \( \mathcal{F}(\mathbf{p}) \) is the signal value at position \( \mathbf{p} \),
    \item \( d\mathbf{p} \) is the infinitesimal path element.
\end{itemize}
To reduce computational cost, we can use nearest neighbors method to select the points for the IDW computation.

\begin{table*}[!t]
\centering
\caption{Comparison of reconstruction results for different images using the proposed method and randomly choosing method. The values are presented as PSNR (dB) / SSIM.}
\begin{tabular}{|c|c|c|c|c|c|c|c|}
\hline
\textbf{Image} & \textbf{Name} & \textbf{Size} & \textbf{Total Samples} & \textbf{Sampling (\%)} & \textbf{Initial } & \textbf{Random} & \textbf{Proposed}\\ \hline
Image 1 &  Synthetic Shapes Image & \( 256 \times 256 \)  & 262139 & 21.4\% & 7.3/0.62 & 15.1/0.74 & \textbf{23.7/0.91}  \\ \hline
Image 2 & Phantom & \( 256 \times 256 \) & 262139 & 16.2\% & 12.8/0.70 & 17.6/0.82 & \textbf{22.4/0.88}  \\ \hline
Image 3 & Cameraman & \( 256 \times 256 \)  & 262139 & 22.4\% & 18.6/0.63 & 24.9/0.71 & \textbf{30.1/0.87}  \\ \hline
Image 4 & WSe2 nanoflake  & \( 401 \times 81 \) & 129921 & 23.5\% & 16.9/0.52 & 21.3/0.69 & \textbf{28.3/0.85}  \\ \hline
\end{tabular}
\label{table:results}
\end{table*}

\section{Algorithm}
We have discussed all the steps for a single iteration. The complete algorithm is summarized in Algorithm \ref{alg:fly-sanc}.

\begin{algorithm}[h!!!]
\caption{Iterative Adaptive Fly-Scan Completion Algorithm
}
\label{alg:fly-sanc}
\begin{algorithmic}[1]
\REQUIRE Initial fly-scan $f^0$ and anchor points, number of anchor points $N$, uncertainty scaling factor $\alpha$, characteristic length scale $\ell$, number of optimization iterations $S$ (or other stopping condition), number of scan iterations $K$ (or other stopping condition)\\
    \STATE Initialize $f^0, N, \alpha, \ell, S, K$.\\
    \STATE Set $k=0$.\\
   \WHILE{$k < K$}   
   \STATE Compute $\Omega_k^0$ by \eqref{eq:score}.
   \STATE Compute $\Omega_k^s$ by \eqref{eq:ADAM} \textbf{for} $0$  \textbf{to} $S$.
    \STATE Compute $\Gamma_k^*$ by \eqref{eq:huristic}.
    \STATE Obtain $\mathcal{F}(\Gamma_k^*)$ by fly-scan.
    \STATE Compute $f^{k+1}$ by \eqref{eq:IDW}.
    \STATE $k \coloneqq k+1$.
   \ENDWHILE
   \RETURN Recovered image $\hat{f}=f^{K}$.\\
\end{algorithmic}
\end{algorithm}

\section{Experiment}
\label{sec:Experiment}

\begin{figure*}[hb]
\centering
\begin{tabular}{ccccccc}
\hspace{-.26cm} 
\includegraphics[width=3.4cm]{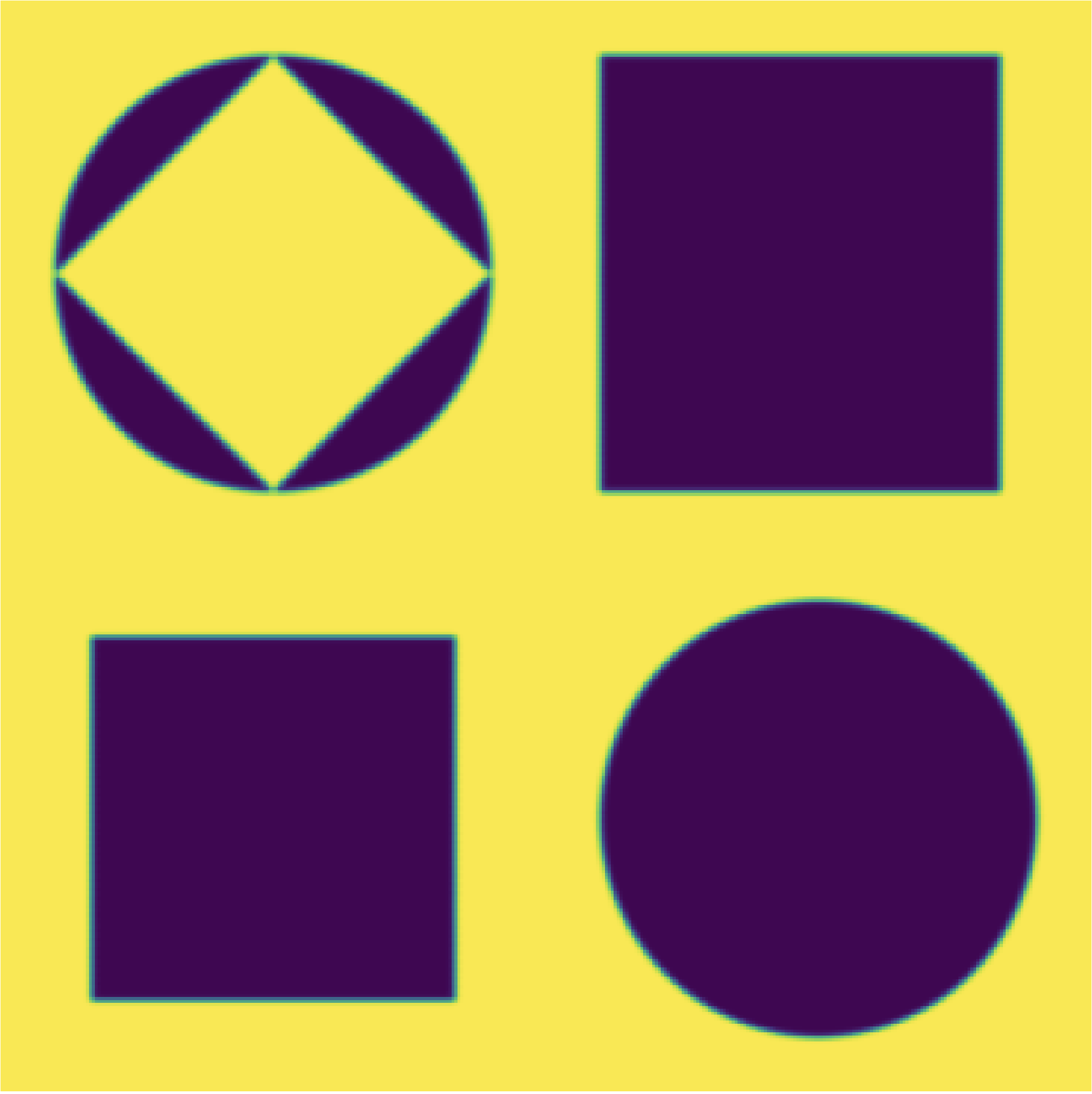} 
\hspace{-.22cm} 
&
\hspace{-.22cm} 
\includegraphics[width=3.4cm]{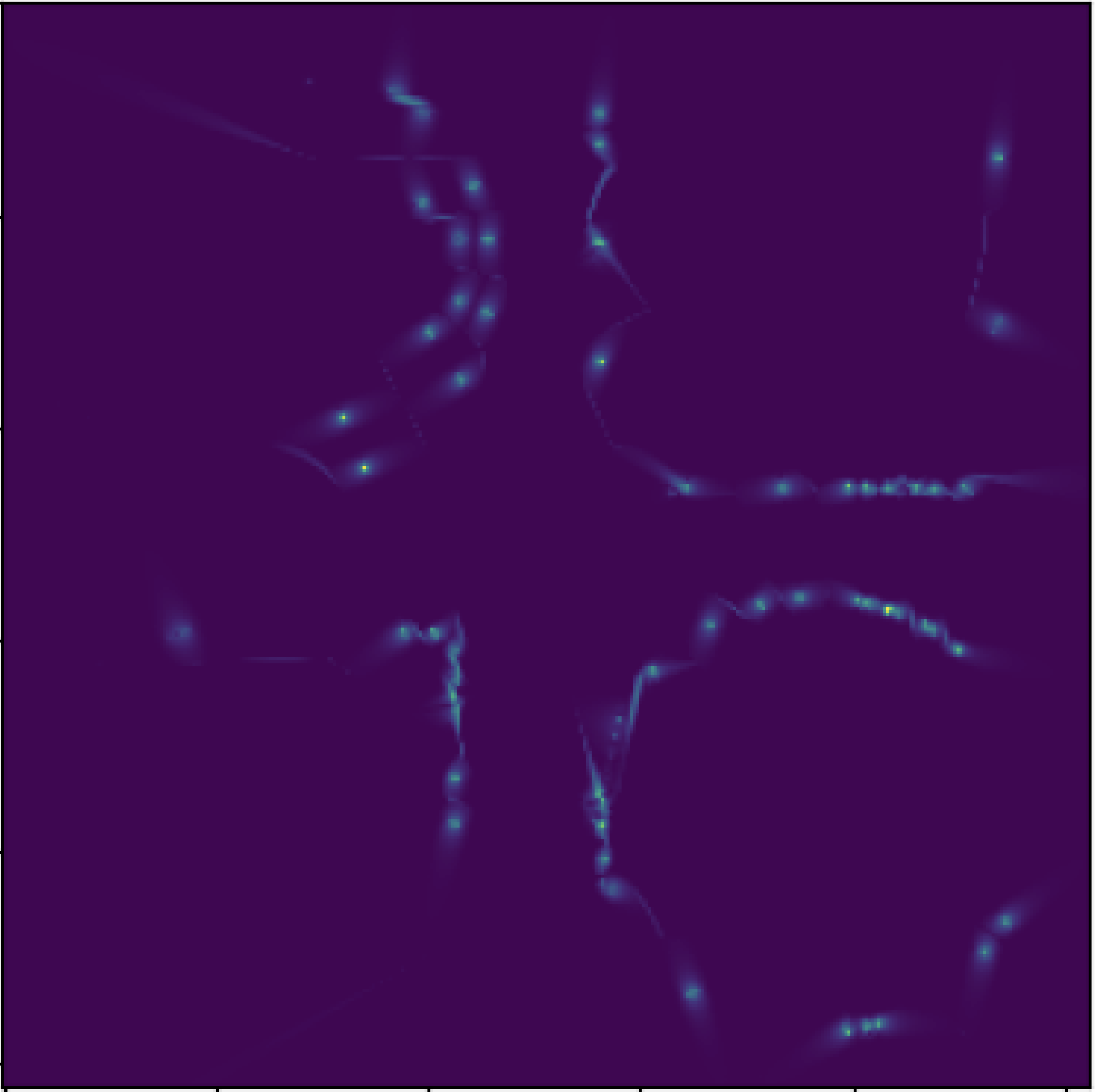} 
\hspace{-.22cm}  
&
\hspace{-.26cm} 
\includegraphics[width=3.4cm]{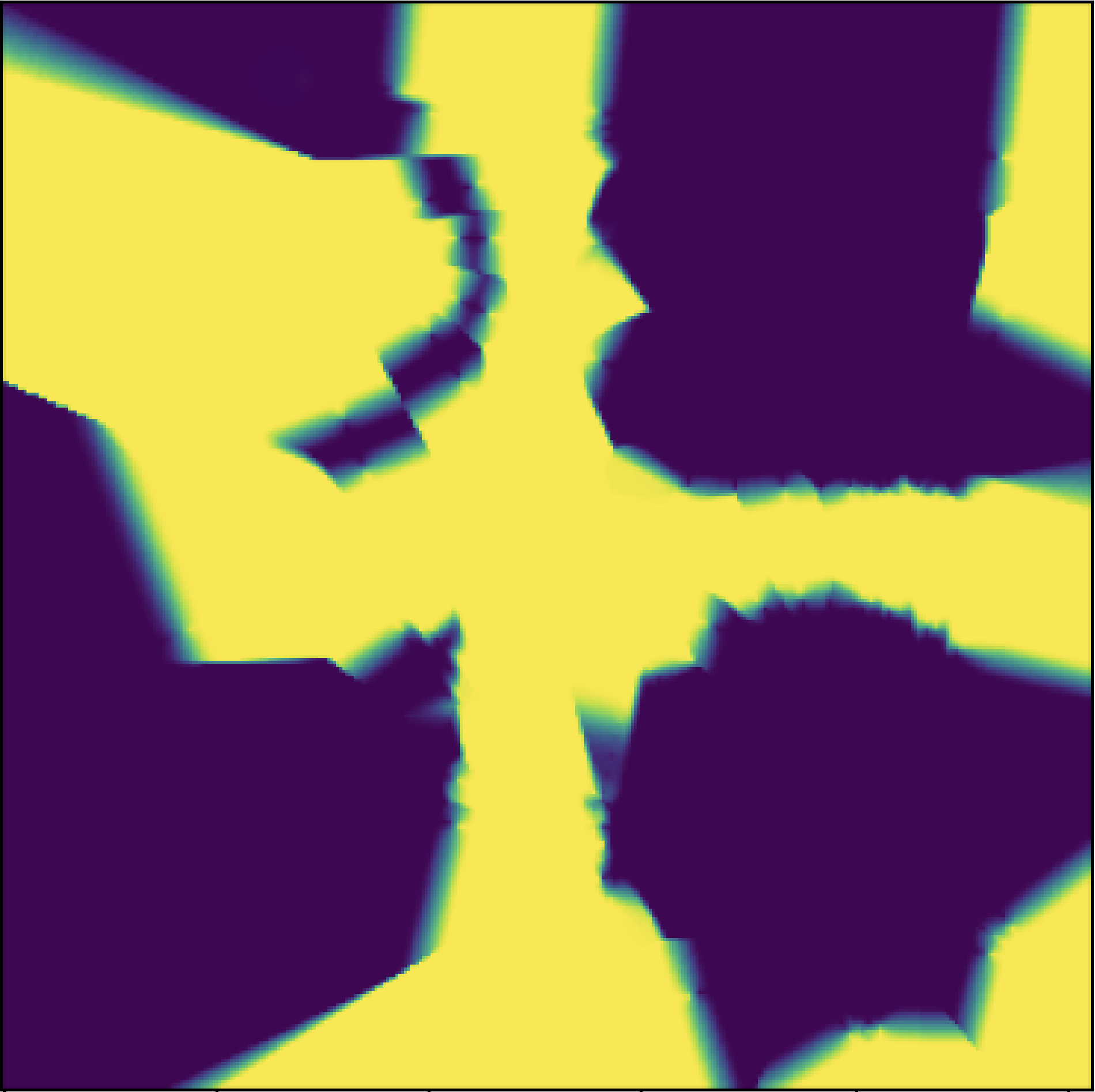} 
\hspace{-.22cm} 
&
\hspace{-.22cm} 
\includegraphics[width=3.4cm]{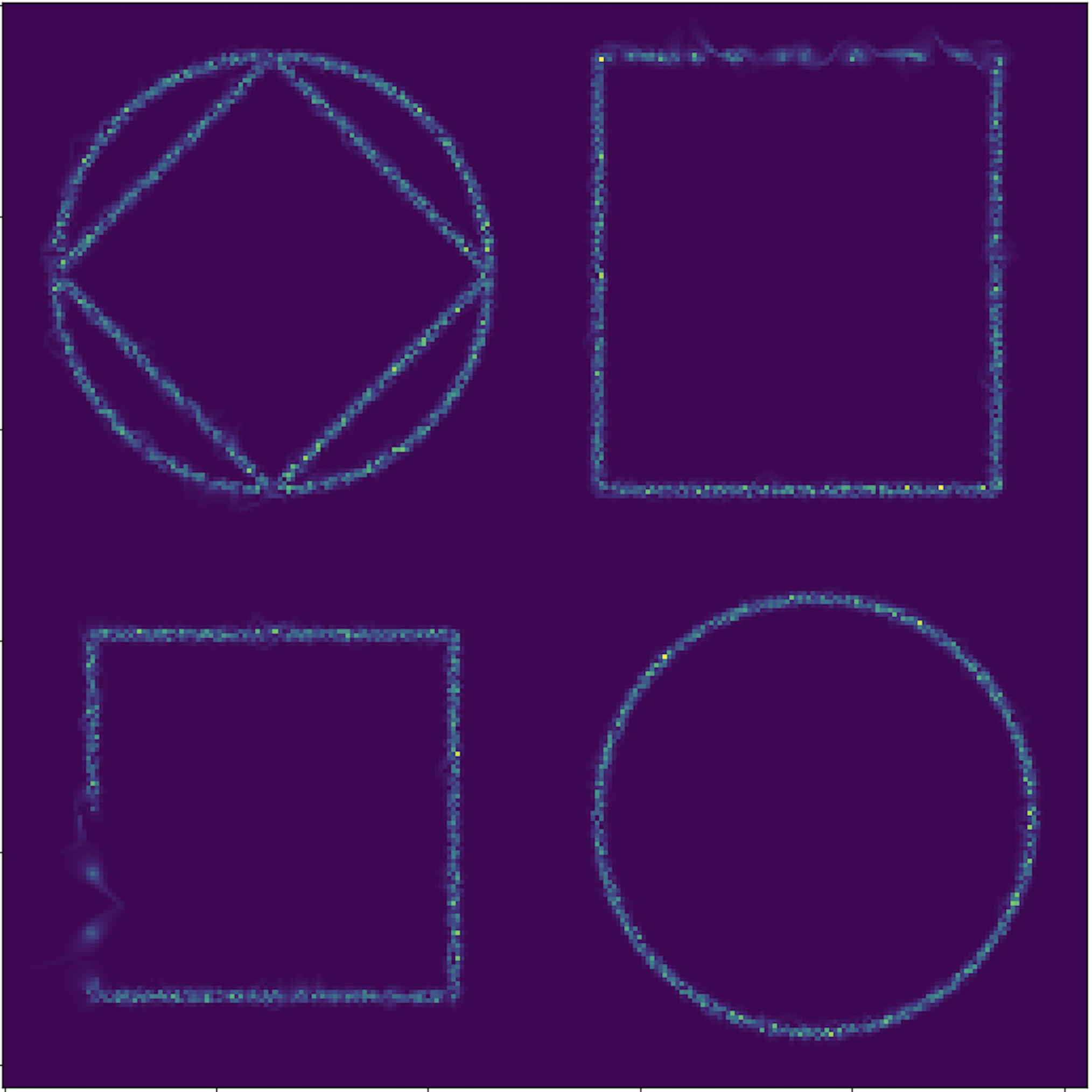} 
\hspace{-.22cm}  
&
\hspace{-.22cm} 
\includegraphics[width=3.4cm]{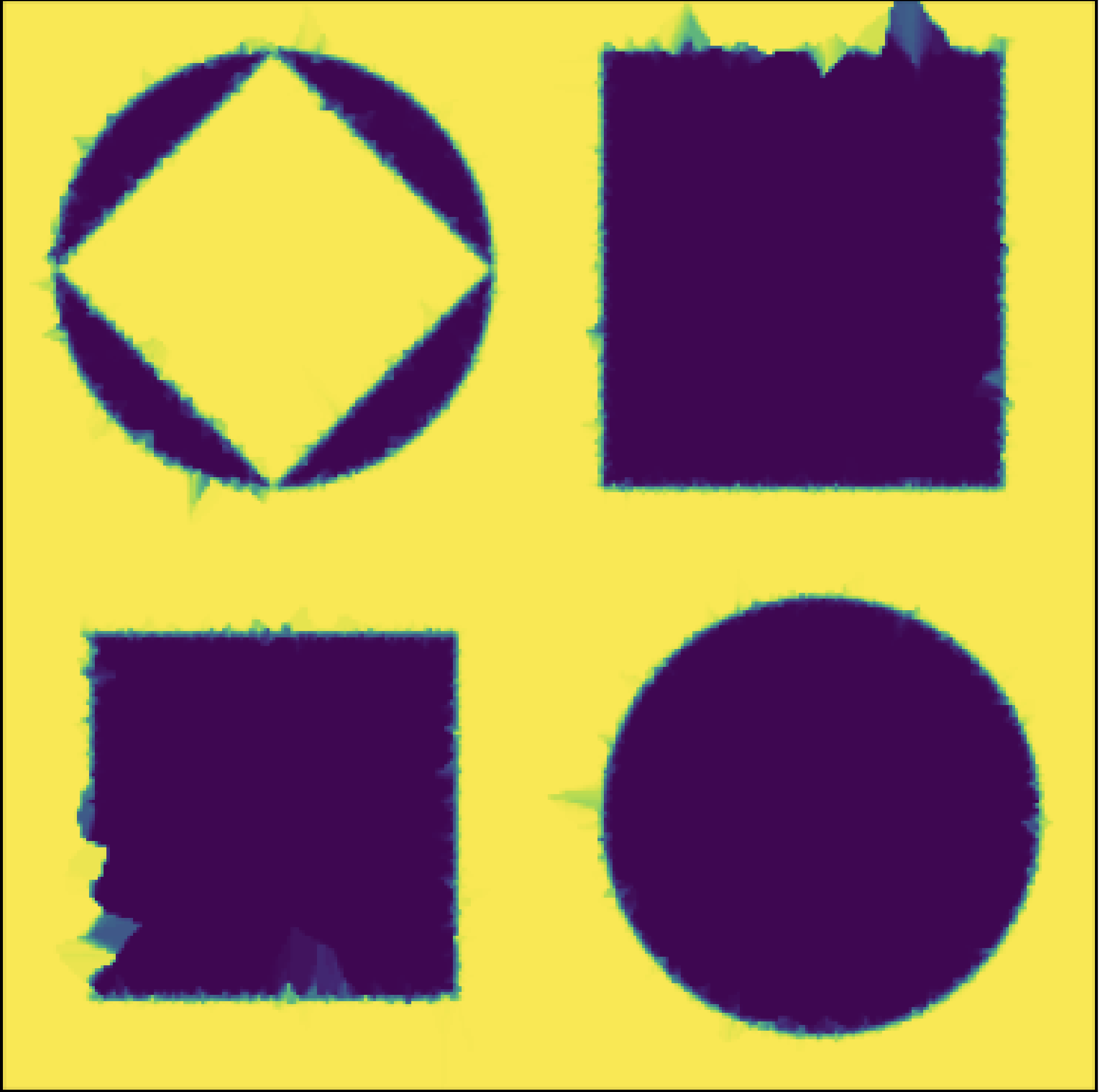} 
\hspace{-.22cm}  
\\
(a) $f^*$ & (b) $\nabla f^0$ & (c) $f^0$  
& (d) $\nabla \hat{f}$ & (e) $\hat{f}$ 
\end{tabular}

\caption{The results for image 1: 
(a) The raster fly-scan result $f^*$.
(b) The initial reconstruction gradient.
(c) The initial IDW reconstruction using $7.1\%$ samples.
(d) The final reconstruction gradient.
(e) The final reconstruction using $21.4\%$ samples.}
\label{fig:img1}
\end{figure*}

To validate our method for image completion in the fly-scan setting, we conduct experiments on four distinct images. The first image is a synthetic shape image from \cite{di2016optimization}, with a resolution of \( 256 \times 256 \), which is used to demonstrate the efficiency of our algorithm in identifying edges. The second and third images are the well-known Shepp-Logan Phantom \cite{shepp1974fourier} and Cameraman images respectively, both with \( 256 \times 256 \) resolution, showcasing the generalization capability of our method. The fourth image is a WSe\(_2\) nanoflake specimen \cite{kandel2023demonstration} with a resolution of \( 401 \times 81 \), representing a typical application scenario in x-ray fly-scan imaging. In addition to the primary experiments, we also evaluate images reconstructed using a uniform random selection of points instead of those obtained by optimizing using our objective function. This comparison highlights the effectiveness of our proposed loss function. Each experiment is conducted independently \( 10 \) times on a 2023 MacBook Air with an M3 chip, and the average result is reported as the final outcome. The primary objectives of these experiments are
\begin{enumerate}
    \item To verify that our proposed objective function can efficiently generate a nearly optimal set of anchor points.
    \item To demonstrate the generalization of our proposed method.
    \item To examine the efficiency of our method in reconstructing images using less than 30\% of the raster scan samples.
\end{enumerate}

For all experiments, the initial fly-scan \( f^0 \) is generated using random $30$ anchor points, with the total number of fly-scan samples comprising less than \( 7\% \) of the raster scan samples. In each iteration, we select \( N = 600 \) anchor points and set the number of iterations to \( K = 16 \) for first three images. For the last image, we choose  \( N = 500 \) and \( K = 12 \) because of its smaller size.  The uncertainty scaling factor \( \alpha \) is chosen as \( 10 \) and the characteristic length scale \( \ell \) is set to \( 4 \) in the objective function. The optimization iteration count is set to \( S = 50 \). Additionally, we configure our fly-scan setting to closely mimic real-world technical conditions. Each pixel in the image corresponds to \( 1 \) nanometer, and the scanning speed is $1$ nanometer per second. The exposure time is \( 0.5 \) seconds, and the dead time between scans is \( 0.02 \) seconds.

Finally, we employ two common and widely used metrics. The first is the peak signal-to-noise ratio (PSNR) \cite{hore2010image}, and it is defined as
\begin{align}
\text{PSNR}(\hat{f}, f^*) = 10 \cdot \log_{10} \left( \frac{\mathrm{MAX}^2}{\text{MSE}(\hat{f}, f^*)} \right),
\end{align}
where $\mathrm{MAX}$ is the maximum possible pixel value of the image, and $\text{MSE}(\hat{f}, f^*) = \frac{1}{n} \sum_{i=1}^{n} \left( \hat{f}_i - f^*_i \right)^2$ is the mean squared error between the reconstructed image $\hat{f}$ and the ground truth $f^*$. We also compute the structural similarity index measure (SSIM) \cite{hore2010image, nilsson2020understanding} for each experiment, which is defined as
\begin{align}
\text{SSIM}(\hat{f}, f^*) = \frac{(2\mu_{\hat{f}} \mu_{f^*} + C_1)(2\sigma_{\hat{f}f^*} + C_2)}{(\mu_{\hat{f}}^2 + \mu_{f^*}^2 + C_1)(\sigma_{\hat{f}}^2 + \sigma_{f^*}^2 + C_2)},
\end{align}
where $\mu_{\hat{f}}$ and $\mu_{f^*}$ are the means of the reconstructed image $\hat{f}$ and the ground truth image $f^*$, $\sigma_{\hat{f}}^2$ and $\sigma_{f^*}^2$ are the variances, $\sigma_{\hat{f}f^*}$ is the covariance, and $C_1, C_2$ are small constants to stabilize the division. SSIM captures human visual perception of image quality and is sensitive to structural preservation.
Since our goal is to reconstruct the image generated by the fly-scan, we compare our image completion results to the raster fly-scan results, not the real ground truth. The final results are summarized in Table~\ref{table:results}.

\subsection{Experiment 1}
In this experiment, we apply the proposed adaptive fly-scan approach to a synthetic image that includes four distinct shapes, as shown in Fig.~\ref{fig:img1}. This image was first published in \cite{di2016optimization} and additional details about the image can be found in the cited paper. The initial reconstruction is highly deformed, making it very difficult to distinguish the shapes. However, the results of our proposed approach yield a much cleaner reconstruction. From Table~\ref{table:results}, our method uses only \( 21.4\% \) of the total samples while achieving a PSNR of \( 23.7 \, \mathrm{dB} \). In comparison, using the same number of samples selected at random achieves only \( 15.1 \, \mathrm{dB} \). The key to the higher PSNR achieved by our proposed method lies in the objective function and score function, which effectively locate positions more likely to contain critical information.

\subsection{Experiment 2}
\begin{figure}[t]
\subfigure[$f^*$ of Phantom]{\label{fig:a}\includegraphics[width=0.3\linewidth]{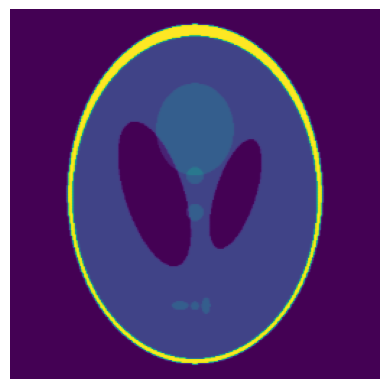}}
\subfigure[$f^0$ of Phantom]{\label{fig:b}\includegraphics[width=0.3\linewidth]{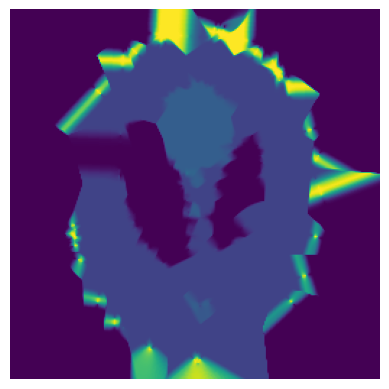}}
\subfigure[$\hat{f}$ of Phantom]{\label{fig:c}\includegraphics[width=0.3\linewidth]{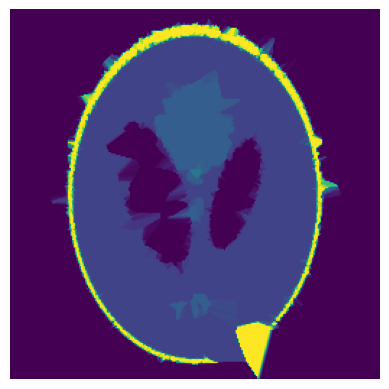}}\\
\subfigure[$f^*$ of Cameraman]{\label{fig:d}\includegraphics[width=0.3\linewidth]{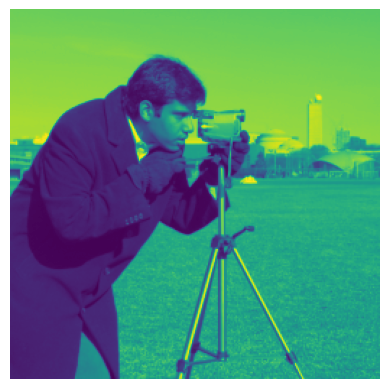}}
\subfigure[$f^0$ of Cameraman]{\label{fig:e}\includegraphics[width=0.3\linewidth]{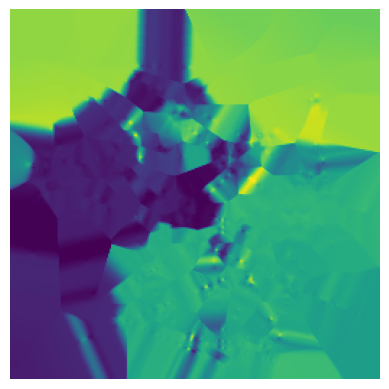}}
\subfigure[$\hat{f}$ of Cameraman]{\label{fig:f}\includegraphics[width=0.3\linewidth]{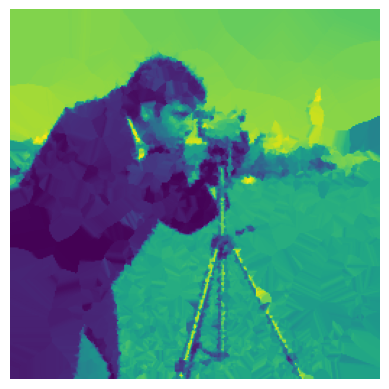}}
\caption{The first row shows the results for Image 2, while the second row shows the results for Image 3. From left to right, the columns represent the raster fly-scan result, the initial result, and the final reconstruction.}
\label{fig:img23}
\end{figure}

This experiment primarily aims to demonstrate that the proposed framework can generalize to commonly used images. We choose the Phantom and Cameraman images as two examples, which are widely used in the field of image reconstruction. The visualized results are shown in Fig.~\ref{fig:img23}. 

For the Phantom image, we use only \( 16.2\% \) of the total samples while achieving a PSNR of \( 22.4 \, \mathrm{dB} \) in the final results. Referring to Table~\ref{table:results}, the reader may notice that this image uses a smaller sampling percentage compared to other images. The key reason is that we select samples based on anchor points generated by our proposed objective and score functions. To demonstrate the robustness of our method, we fixed the anchor points number in each iteration and initial parameter settings for all images. Consequently, the optimal anchor points set used for reconstruction quickly converges to the regions of interest determined by the objective function. Since the edges in the Phantom image form an oval, many anchor points are concentrated in that region [see  in Fig.~\ref{fig:img23}(c)]. While relaxing the score and objective functions could further improve the results, this experiment shows that our proposed method still performs very well even for special cases like the Phantom image.

The Cameraman image, being more complex and detail-rich, achieves a very high PSNR using our framework. Though utilizing only \( 22.4\% \) of the raster scan samples, yet achieves a PSNR of \( 30.1 \, \mathrm{dB} \) in the final results. Many details are successfully recovered, such as the buildings in the background, further highlighting the effectiveness of the proposed approach. Both results demonstrate the strong generalization capability of the proposed framework.open

\subsection{Experiment 3}\label{EX3}

\begin{figure}[h]
\subfigure[Ground Truth]
{\label{fig:i}\includegraphics[width=1\linewidth]{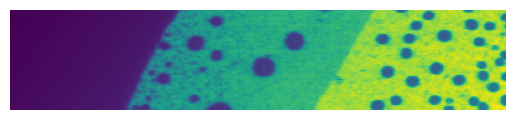}}\\
\subfigure[Initial Scan]
{\label{fig:j}\includegraphics[width=1\linewidth]{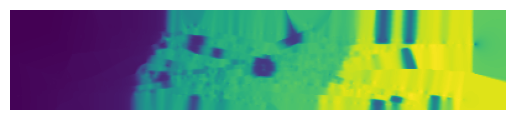}}\\
\subfigure[Final Scan]
{\label{fig:k}\includegraphics[width=0.97\linewidth]{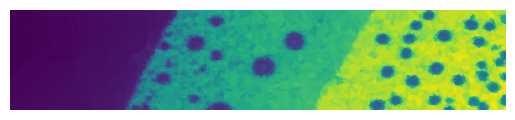}}
\caption{Results for Image 4: The first row shows the raster fly-scan result, the second row shows the initial result, and the third row shows the final reconstruction.}
\label{fig:img4}
\end{figure}

In the final experiment, we apply our proposed method to an image of a WSe$_2$ nanoflake specimen, and the visualized results are presented in Fig.~\ref{fig:img4}. The image was originally acquired using a scanning x-ray microscopy setup in step-scan mode at the Advanced Photon Source and was first published in \cite{kandel2023demonstration}; the raw data is available at \cite{kandel2023zenodo}. More details about the preparation and objectives of the image can be found in the cited paper. This experiment aims to demonstrate that our proposed approach performs very well on a typical fly-scan application image. Using \( 23.5\% \) of all samples, the fly-scan approach achieves a PSNR of \( 28.3\, \mathrm{dB} \), successfully recovering many features, such as holes of different sizes.

\begin{figure}[H]
\subfigure[Raster Scan Path]
{\label{fig:yy}\includegraphics[width=1\linewidth]{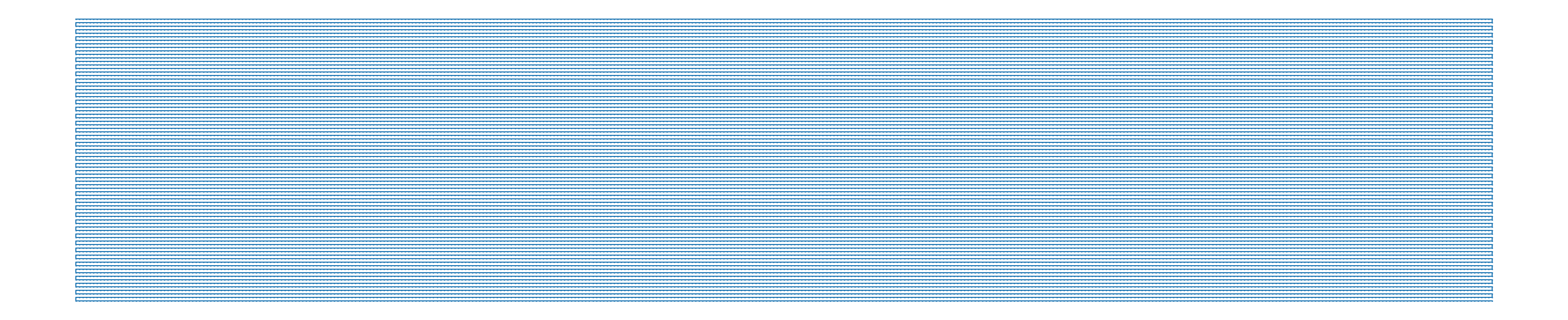}}\\
\subfigure[Adaptive Scan Path]
{\label{fig:y}\includegraphics[width=1\linewidth]{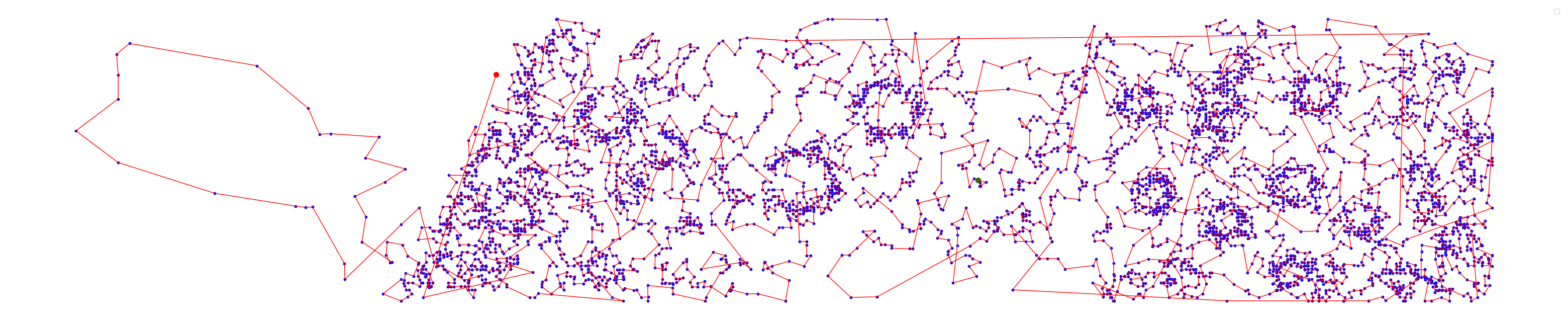}}\\
\subfigure[Gradient Map]{\label{fig:z}\includegraphics[width=0.97\linewidth]{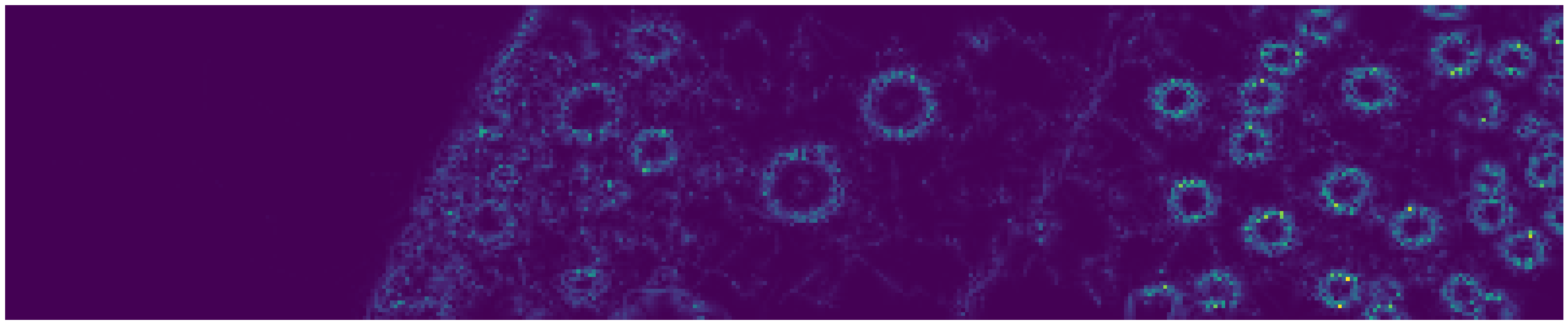}}
\caption{Plot (a) shows the raster scan path, Plot (b) shows the scan path generated by our proposed approach, and Plot (c) shows the gradient of the final reconstruction.}
\label{fig:img4_path}
\end{figure}

Additionally, we provide the scan paths for both the raster fly-scan and our proposed adaptive fly-scan (see Fig.~\ref{fig:img4_path}). Our proposed scan path is very "smart" as it avoids wasting time in uniform areas and focuses on edges in the image, which are weighted more by the objective function. The last row in the plot shows the gradient of the final reconstruction, which closely resembles the ground truth.

\subsection{Parameter Analysis}
In this section, we investigate the impact of key parameters on the performance of the proposed framework. Our framework includes two types of parameters. The first category consists of size-related parameters:
\begin{itemize}
    \item the number of anchor points $N$ in Eq.~\eqref{eq:EWUF},
    \item the size of the candidate subset $\chi$ in Eq.~\eqref{eq:HA},
    \item the number of samples used in the image completion subproblem in Eq.~\eqref{eq:IDW}.
\end{itemize}
The effects of these parameters are straightforward: increasing their values generally improves accuracy and enhances the framework’s performance. However, larger sizes will result in higher computational costs and slower runtimes.

The second category is hyperparameters. Specifically, we analyze the effects of
\begin{itemize}
    \item Uncertainty scaling factor $\alpha$: it controls whether the objective function in Eq.~\eqref{EQ:AC} focuses more on uncertain regions or on regions with rapidly changing features,
    \item Characteristic length scale $\ell$: it controls the spatial sensitivity of the EWUF by determining how strongly nearby regions influence each anchor point's uncertainty.
\end{itemize}

For simplicity, we fix the noise standard deviation $\sigma = 1$ throughout all experiments, as we consider the noise-free situation and the trade-off between uncertainty and gradient terms is already controlled by the hyperparameter $\alpha$. To analyze these parameters, we examine Image 4 from experiment 3. During the search process, we test the algorithm on all combinations of potential parameter values to find the optimal PSNR. To clearly illustrate the sensitivity of each parameter in this section, we fix one parameter while testing the another one.

\begin{figure}[H]
\centering
\includegraphics[width=0.48\textwidth]{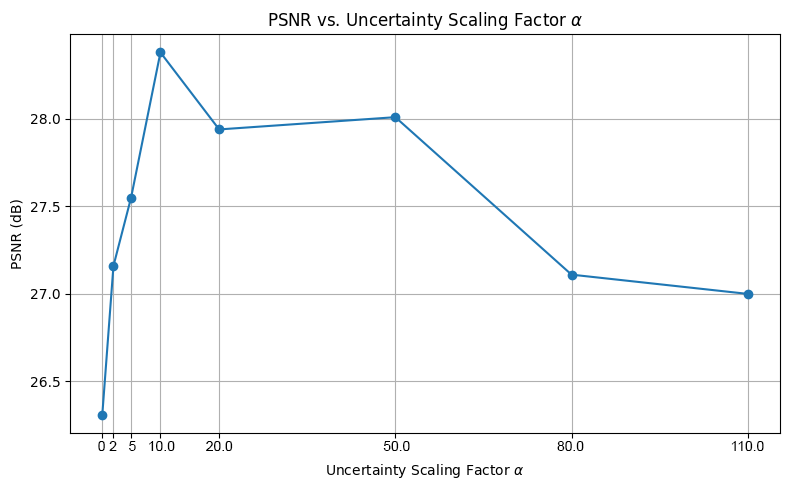}
\caption{PSNR performance of the proposed method with varying $\alpha$ values for Image 4.}
\label{fig:PA11}
\end{figure}

First, we analyze the parameter $\alpha$ while fixing $\ell = 4$. Figure~\ref{fig:PA11} shows the PSNR of the final scanned image with $\alpha \in \{$0.1, 2, 5, 10, 20, 50, 80, 110$\}$. A larger $\alpha$ emphasizes exploration by prioritizing uncertain regions, while a smaller $\alpha$ focuses more on exploitation of high-gradient regions. It can be seen that there is an optimal $\alpha$ for Image 4; however, the performance variation across different $\alpha$ values is limited. The maximum PSNR difference is approximately 2 dB, corresponding to less than an $8\%$ variation, suggesting that our framework is relatively stable with respect to $\alpha$.

\begin{figure}[H]
\centering
\includegraphics[width=0.48\textwidth]{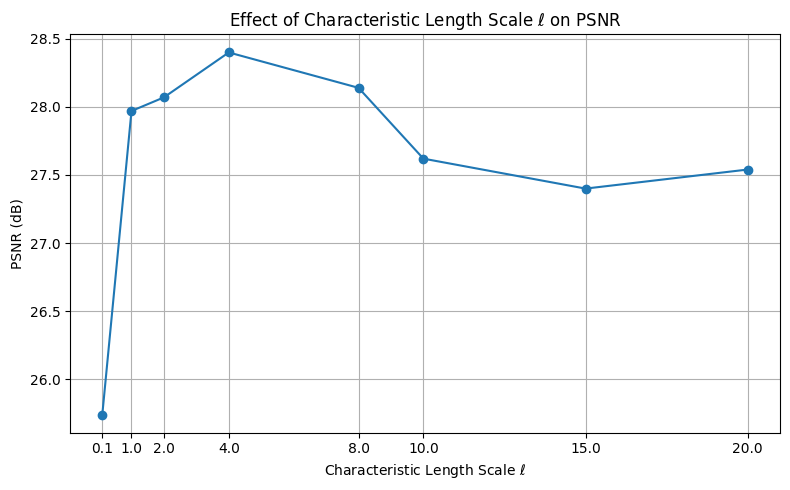}
\caption{PSNR performance of the proposed method with varying $\ell$ values for Image 4.}
\label{fig:PA2}
\end{figure}

Second, we analyze the parameter $\ell$ while fixing $\alpha = 10$. Figure~\ref{fig:PA2} shows the PSNR of the final scanned image with $\ell \in \{$0.1, 1, 2, 4, 8, 10, 15, 20$\}$. A larger $\ell$ emphasizes local details by focusing on nearby neighbors and encourages new anchor points to be placed in fine-scale regions because the large $\ell$ means the Gaussian kernel, which estimates the correlation, drops more slowly with distance, so the correlation tends to be overestimated and the uncertainty tends to be underestimated, and that leads the model to focus more on exploitation rather than exploration. In contrast, a smaller $\ell$ leads to more global averaging across a wider neighborhood, causing anchor points to be more spread out as the uncertainty gradients become smoother, promoting exploration of broader regions.

The observed PSNR variation is smaller than expected, especially for large $\ell$. One key reason is that we are already using $23.5\%$ of the total points, and the image does not have significant variation across local regions. Additionally, even when considering the extreme case of $\ell = 0.1$, the maximum PSNR difference is less than 2.5 dB, further demonstrating that our framework is highly stable with respect to $\ell$.

\section{Discussion}
In this section, we discuss the limitations of the proposed framework and potential future research directions.

\subsection{Limitations}

\subsubsection{Initial Path}
In the proposed framework, the initial scan path is randomly generated and utilizes approximately $6\%$ of the total samples. This approach performs well across all our experiments, suggesting that it is effective for most images. However, for certain special cases where important features are concentrated in a small region of the image, the random initial scan may capture very little useful information. This can lead to anchor points being selected in a less informed manner, resulting in inefficient scanning.

One way to address this limitation is by incorporating prior information about specific image types and designing a more tailored initial scan strategy. Some studies have explored more efficient initial scan methods \cite{stefos1989effect}, but applying these techniques into a fly-scan framework could remain a challenge.

\subsubsection{Final Path}
The proposed framework generates an efficient scan path for fly-scan applications. However, if we take a closer inspection of the final scan path in Fig.~\ref{fig:z}, we may observe that some path segments exhibit sharp angles. In real-world experiments, the motor may not be able to follow these sharp turns smoothly. This issue arises because the proposed framework focuses solely on path efficiency. A potential solution is to add a penalty or regularization term to the objective function, taking hardware constraints into account, to encourage smoother scan paths.

\subsubsection{Noisy Images}
To the best of our knowledge, this paper is the first to investigate adaptive fly-scan path generation. Therefore, we focus on designing a general framework applicable to standard scenarios. Consequently, all images used in our three experiments are assumed to be noise-free. However, in real experimental environments, unavoidable noise may be introduced, including Gaussian noise \cite{luisier2010image, wink2004denoising} and low-photon noise, such as Poisson noise \cite{rodrigues2008denoising, willett2010poisson} or negative binomial noise \cite{lu2024sparse}.

Our proposed framework does include some inherent noise reduction. First, the fly-scan method itself provides slight smoothing, as measurements are integrated over a dead time, reducing the influence of high-frequency noise. Second, previous studies suggest that the IDW algorithm can help reduce noise \cite{chen2016denoising} by smoothing isolated coordinates. However, when interpolated values are derived from a noisy scan in the previous iteration, the IDW algorithm may introduce issues. For example, it may assign a noisy measurement to one coordinate while a nearby coordinate has already been scanned and contains a value closer the signal. Even if these two coordinates share the same value in the original image, after interpolation they may exhibit a significant difference, resulting in a large gradient. This misleading gradient could negatively impact the proposed framework by causing the subsequent scan path to be less efficient.

Although many denoising algorithms have been proposed in previous studies, integrating these denoising modules into the proposed adaptive fly-scan framework remains a challenge. One reason is the lack of prior information about the experimental environment, making it difficult to identify the type of noise. Another reason is that denoising could result in the loss of true scanned information, which is particularly critical in adaptive fly-scan and image completion tasks compared to normal denoise tasks.

\subsection{Future Research}
Based on the limitations of our proposed framework and recent developments in related research, we suggest the following future research directions:

\begin{itemize}
    \item \textbf{Self-adaptive fly-scan:} Since the random initial scan may not be effective in certain special cases, such as featureless images, a self-adaptive strategy could be explored. For instance, a verification layer could be added after the initial scan. If the average gradient of the initially scanned anchor points is very small, which indicates limited useful information was obtained, an additional initial scan could be performed in a different region. Furthermore, to address sharp turn angles in the final scan path, which may be impractical for hardware, penalty terms incorporating hardware constraints or a simple smoothing check function could be added to the point selection process to ensure more feasible paths.
    
    \item \textbf{Adaptive fly-scan with denoising:} As discussed in the previous subsection, incorporating a denoising mechanism could enhance the performance of the proposed framework. A comprehensive study investigating different denoising strategies would be valuable. Potential directions include adding a diagonal noise matrix to the covariance matrix \cite{carlson1988covariance, pascal2008performance}, introducing denoising regularization terms (such as total variation \cite{vogel1996iterative, chen2010adaptive}) to mitigate IDW-related issues, or integrating a dedicated denoising module before each scan iteration or after each image completion step.
    
    \item \textbf{Deep learning-based adaptive fly-scan:} In the proposed framework, the scan path is obtained by optimizing an objective function that relies on uncertainty (exploration) and gradient (exploitation) information. However, there may exist additional important features for path selection which are difficult to explicitly define by hand. Using deep learning models, such as convolutional neural networks (CNNs) or Transformers \cite{vaswani2017attention, dosovitskiy2020image}, could help extract lower-level features from the images and sampling density maps to guide the scan path. The key motivation is that deep models can learn complex spatial and contextual patterns, enabling a more expressive and adaptive acquisition strategy. In the context of fly-scan, both end-to-end architectures (where the model directly predicts the anchor points or paths from reconstructed images) or unrolling methods \cite{monga2021algorithm} combined with the objective function (where neural networks replace certain operators in the iterative minimization of the objective) could be explored for future research.

    \item \textbf{High-dimensional adaptive fly-scan:} The fly-scan technique can be extended to 3D imaging modalities, such as 3D ptycho-tomography. Designing a fully adaptive fly-scan path in high-dimensional spaces could provide better efficiency compared to an approach that combines multiple 2D fly-scans from different orientations.
\end{itemize}

\section{Conclusion}
In this paper, we proposed an adaptive fly-scan framework designed to enhance the efficiency of fly-scan imaging. To the best of our knowledge, this paper is the first work in the adaptive path reseach for the fly-scan technique. By focusing the scanning process on regions of interest and employing differentiable image completion techniques to reconstruct unscanned areas, our approach addresses the inefficiencies of traditional raster fly-scan methods. The proposed framework integrates an efficient score function and objective function to dynamically generate optimal anchor points, enabling precise and adaptive scanning paths in each iteration.

Through three experiments on four images, including synthetic and real-world images, we demonstrated that our framework consistently achieves high-quality reconstructions while utilizing fewer than $30\%$ of the measurements made by dense raster scans. Furthermore, our method generalizes well to different imaging scenarios, adapting to the unique structures of complex and detail-rich images. Our work provides a stable approach for more effective and resource-efficient imaging systems.

\section{Acknowledge}
This research is based on work supported by Laboratory Directed Research and Development (LDRD) funding from Argonne National Laboratory, provided by the Director, Office of Science, of the U.S.~DOE under Contract No.~DE-AC02-06CH11357. The research was also supported by the Applied Mathematics activity within the U.S.~Department of Energy, Office of Science, Advanced Scientific Computing Research and Basic Energy Science, under Contract DE-AC02-06CH11357. We thank Dr.~Saugat Kandel for sharing the scanning x-ray microscope data used in the demonstration of this work. We also thank Dr.~Yanqi (Grace) Luo and Dr.~Sarah Wieghold for helpful discussions.



\bibliographystyle{IEEEbib}
\bibliography{icme2025arxiv}

\end{document}